\documentclass{article}
\usepackage[dvips]{graphicx}

\begin{document}

\begin{center}
Talk given at the 2K1BC Workshop\\
 {\it Experimental Cosmology at Millimetre Wavelengths}\\
 July 9-12, 2001, Breuil-Cervinia (Italy)
\end{center}
\vskip2mm
\hrule
\vskip1cm

\begin{center}
\huge{\bf Primordial molecules at millimeter wavelengths}
\vskip4mm
\large{\bf D. Puy$^\dagger$, M. Signore$^\ast$}
\vskip4mm
{\small
$^\dagger$Institute of Theoretical Physics, Z\"urich and PSI-Villigen 
(Switzerland), email: puy@physik.unizh.ch
\vskip1.5mm
$^\ast$Observatoire de Paris-DEMIRM, Paris (France), 
email: monique.signore@obspm.fr}
\end{center}
\vskip9mm
\noindent
{\bf Abstract.}
Chemistry plays a particular role in astrophysics. After atomic hydrogen, 
helium and their ions, the Universe probably contains more mass in molecules 
than in any other species. Molecule formation in the early, pre-galactic 
Universe may have had much to do with the formation of galaxies themselves. 
In this context the possible interaction between primordial molecules and 
photons of the Cosmic Microwave Background (CMB) is very important through 
the theoretical perspectives and constraints which could give some 
information on the theory of the large scale structure formation. 
\\
In this paper we recall the more recent progresses on the chemistry of the 
early Universe, and describe the importance of molecules in the formation 
phase of proto objects. A special attention is done concerning the {\it case 
of $LiH$}.
\section{Introduction}
Molecules are found in a large variety of astronomical environments. 
They are now widely used as diagnostic probes of the physical conditions 
in which they occur. The diversity of molecular environments has helped 
to stimulate interest in a variety of different chemical processes. 
\\
According\footnote{see the talk of Signore-Puy in this proceedings 
and Signore \& Puy \cite{Signore:1}.} to the standard Big Bang 
cosmology, the space of the Universe expands adiabatically, cooling from an 
extreme initial temperature and density. Thus at about one second after 
the Big Bang, the temperature of the Universe remains hot and around 
10$^{10}$ K. At this stage, the collisions between neutrons and protons can 
form deuterium, and open the way of the primordial 
nucleosynthesis through other fusion reactions (see Signore \& Puy 
\cite{Signore:1} and references therein). 
At about 100 sec the 
nucleosynthesis epoch is over, thus most neutrons are in $^4He$ nuclei, and 
most protons remain free while smaller amounts of $D$, $^3He$ and $^7Li$ 
are synthetized. The low densities, the Coulomb barriers and stability gaps 
at masses 5 and 8 worked against the formation of heavier elements.  
\\
After the nucleosynthesis period atoms form by 
recombination of these primordial nucleus with free electrons, leading to 
the thermal decoupling between the matter and the radiation (see Puy-Signore 
\cite{Puy:2001}). The radiative recombination processes were then not reversed 
by photoionization and electron impact ionization, as they had been earlier 
because the supply of energetic photons and electrons had diminished. The 
Universe was transformed into a neutral state, apart from the few relict 
ions and electrons left behind in the expansion.
\\
The chemistry of the early Universe is the chemistry of the elements $H$, 
its isotope D, helium and their isotopic forms. The ongoing physical reactions 
are immense after the recombination of hydrogen, the main processes are 
collisional (ionization, radiative recombination, attachment...) and radiative 
due to the presence of the CMB (photoionization, 
photodetachment...). During the last decade, a large litterature has been 
developped on the chemical networks of the primordial chemistry (Lepp \& Shull 
\cite{Lepp:1984}, Puy et al. \cite{Puy:1993}, Stancil et al. 
\cite{Stancil:1998}, Galli \& Palla \cite{Galli:1998}, Puy \& Signore 
\cite{Puy:1999} and \cite{Puy:2001} for historical description).
 Thus primordial molecules such as $H_2$, $HD$ and $LiH$ formed
\\
The existence of a significant abundance of molecules can be crucial on the 
dynamical evolution of collapsing objects. Because the cloud temperature 
increases with contraction, a cooling mechanism can be important for the 
structure formation, by lowering pressure opposing gravity. This is 
particularly true for the first generations of objects.
\\ 
Thus, in the first part of this communication, we recall the main and recent 
results of the molecular influence on the formation of the proto-objects, 
particularly that of $H_2$ and $HD$ molecules. Interactions between 
primordial molecules and CMB could be important. In the second 
part, we will recall the potential importance of 
$LiH$ molecules on the CMB anisotropies. In particular, the scattering 
process of CMB photons on $LiH$ molecules could play an important role and 
lead to produce secondary anisotropies on the spectrum of CMB. We will 
conclude this communication on the possible outlooks.
\section{Importance of $H_2$ and $HD$ for cosmology}
As we have seen, early Universe chemistry has been previously investigated 
by many authors. $H_2$ and $HD$ molecules are the most 
abundant molecules, and could play a non-negligible role on the formation 
of the first objects of the Universe. These molecules could contribute to 
cooling function and lead to dynamical influence on the collapse mechanism. 
Moreover although the abundance of $H_2$ molecules is rather insensitive 
to the choice of cosmological model, the abundance of $HD$ molecules 
shows large variations.
\subsection{$H_2$ MOLECULE}
Eddington \cite{Eddington:1937}, then Str\"omgren \cite{Stromgren:1939} 
were the firsts to suggest that $H_2$ might exist in the interstellar 
space. Herzberg \cite{Herzberg:1950} described the quantum mechanics of 
homonuclear molecules in some details, which opened important 
theoretical works on $H_2$ in astrophysics. Although the medium is free 
of grains after the cosmological recombination, the formation of $H_2$ comes 
into play through the ions $H^-$ and $H_2^+$. In the post-recombination 
medium the radiation is hotter than the matter. In this context the 
radiative excitation of the rotational levels, which is here more efficient 
than the collisional excitation produces a heating 
-see Puy et al. \cite{Puy:1993}. 
\\
Lepp \& Shull 
\cite{Lepp:1984} were the firsts to point out this important characteristics, 
which was confirmed by Puy et al. \cite{Puy:1993} with a better estimation 
of the thermal function. In the gravitational collapse the 
situation is very different \cite{Puy:1996}; the temperature of CMB is below 
than the 
matter inside of the collapse. Thus the thermal balance between matter and 
radiation leads to produce a molecular cooling function, $H_2$ becoming a 
good coolant agent of the collapse. 
\\
The possibility to observe high redshift systems is mainly associated 
with the presence of quasars, i.e. strong background sources. Most of 
our knowledge of the Universe between $z=1$ and 5 comes from the study 
of the Lyman-$\alpha$ absorbers in the optical range. In a mini-survey for 
molecular hydrogen in eight high-redshift damped Lyman-$\alpha$ systems, 
Petitjean et al. \cite{Petitjean:2000} confirmed the presence of $H_2$ in a 
system toward PKS 1232+082 ($z=2.3377$). They show that there is no 
evidence for any correlation between $H_2$ abundance
\footnote{In this case the upper limits on the molecular fraction 
derived in nine of the systems are in the range $1.2 \times 10^{-7} - 
1.6\times 10^{-5}$.} and relatively heavy element depletion into dust grains. 
\subsection{$HD$ MOLECULE}
The role of deuterium was analyzed by Palla et al. \cite{Palla:1995}, then 
completed by Stancil et al. \cite{Stancil:1998}, in the context of the 
chemical and thermal evolution of the gas component in the post-recombination 
Universe and more recently by Flower \cite{Flower:2000b}. 
Puy \& Signore \cite{Puy:1997} revealed that the $HD$ molecule is the 
main cooling agent, a result which was confirmed later by many authors 
such as Okumurai 
\cite{Okumurai:2000}, Uehara \& Inutsuka \cite{Uehara:2000} and Flower et al. 
\cite{Flower:2000}. Thus $HD$ molecules could have important consequence on 
the problem of fragmentation of primordial clouds in order to form first 
structures like massive stars. Searches for a primordial signature of $HD$ 
is crucial.
\\
One must pointed out that very recently Varshalovich et al. 
\cite{Varshalovich:2001} have analyzed the 
spectrum of the quasar PKS 1232+082 obtained by Petitjean et al. 
\cite{Petitjean:2000}. $HD$ molecular lines have been identified in an 
absorption system at the redshift $z=2.3377$, this is the first detection of 
$HD$ molecules at high redshift.
\section{The case of LiH}
From an initial idea of Zel'dovich, Dubrovich \cite{Dubrovich:1997} showed 
that resonant 
elastic scattering must be considered as the most efficient process in 
coupling matter and radiation at high redshift. He noted that the cross 
section for resonant scattering between cosmic microwave background (CMB) 
photons and molecules is several orders of magnitude larger than that between 
CMB and electrons; even a modest abundance of primordial molecules would 
produce significant Thomson scattering.
\\
During an elastic scattering between CMB photons and primordial molecules, a 
photon is absorbed and reemitted at the same frequency but not in the same 
direction. This process could have negligible effect because of low abundances 
of primordial molecules. Dubrovich \cite{Dubrovich:1993}, Maoli et al. 
\cite{Maoli:1994}, Signore et al. \cite{Signore:1997} showed that this effect 
could alter the primary spatial distribution of the CMB anisotropies. 
More precisely resonant scattering of CMB photons on $LiH$ molecules can be 
particularly efficient for smoothing the primary anisotropies. 
Maoli et al. \cite{Maoli:1994} pointed out that primordial 
molecules such as $LiH$ may play significant role in altering the 
amplitude and power spectrum of CMB anisotropies; the effect depends 
essentially on the $Li$ abundance and the lithium chemistry. They found 
that primary CBR anisotropies may be erased or attenuated for angular 
scales below 10$^o$ and frequencies below 50 GHz, if $LiH$ primordial 
abundance, relative to $H$,  exceed 10$^{-10}$. In 1996 Stancil, Lepp and 
Dalgarno \cite{Stancil:1996} implemented the first complete post-recombination
 lithium chemistry, and concluded that the final abundance of primordial $LiH$
 is below 10$^{-18}$, which ruled out this possibility of erasing the primary 
anisotropies. 
\\
Different studies focused on the chemical evolution in primordial clouds by 
solving a chemical reaction network within idealized collapse models 
(see Puy \& Signore \cite{Puy:1996}, Anninos \& Norman \cite{Anninos:1996}, 
Abel et al. \cite{Abel:2000}). Puy \& Signore \cite{Puy:1997} examined the 
evolution of primordial molecules in a context of gravitational collapse and 
showed that primordial molecules could coexist in a collapsing 
proto-cloud,  particularly during the first phase of gravitational 
collapse. Maoli et al. \cite{Maoli:1996} emphasized the role that elastic 
resonant scattering through $LiH$ molecules can be produced in 
this collapsing structure. 
If the scattering source has a non-zero component of the peculiar velocity 
along the line of sight, they showed that the elastic scattering is no more 
isotropic in the observer frame and molecular secondary anisotropies are 
produced in the CMB spectrum. The angular scale of these secondary 
anisotropies are therefore directly related to the size of 
the primordial clouds. Bougleux \& Galli \cite{Bougleux:1997}, then 
Puy \& Signore \cite{Puy:1998} studied the chemistry of primordial $LiH$ in a 
collapsing protocloud, from the chemical network of Stancil 
et al. \cite{Stancil:1996} and the fully quantum mechanical treatment, of 
the radiative association of the excited $Li$ states, developped by 
Gianturco \& Gori-Giorgi \cite{Gianturco:1996}. We concluded that with this 
chemical network the $LiH$ abundance is closed to $3\times 10^{-18}$, 
leading to very low secondary anisotropies in the CMB. 
\\
Nevertheless a precise analysis of the chemical network shows that most of 
reaction rates are quite uncertain. For example the reaction rate of 
the main reaction which dissociates the $LiH$ molecules:
\begin{equation}
LiH + H \, \rightarrow \, Li + H_2
\end{equation}
is constant and independent of the temperature and of the 
density !

\section{Outlook}
Although astrochemical observations started in the visible, they are 
dominated by the radio and above all by millimitre and sub-millimetre 
observations. Tentative of direct searches of primordial molecules were 
developped this last decade, but the results were not at the level of 
efforts of teams of observers (see for example De Bernardis et al. 
\cite{Debernardis:1993}, Signore et al. \cite{Signore:1994}, Combes \& Wiklind 
\cite{Combes:1998}). Recently Papadopoulos et al. 
\cite{Papadopoulos:2001} revealed the discovery of large amounts of 
low-excitation molecular gas at redshift $z\sim 3.91$. Shibai et al. 
\cite{Shibai:2001} investigated the observability of hydrogen molecules in 
absorption. They argued that the absorption efficiency of the hydrogen 
molecules become comparable with or larger than that of the dust grains in 
the metal-poor condition expected in the early Universe. Thus the absorption 
measurement of the hydrogen molecules could be an important technique to 
explore the primordial gas clouds that are contracting into first-generation 
objects. 
\\
The HERSCHEL satellite \cite{herschel:2001} 
could prove the origins of structure and the chemistry 
at the early interprotostructure medium and furnish a spectral atlas for 
molecules (see Encrenaz et al. \cite{Encrenaz:1997}). A submillimetre spectra 
of protoclouds of gas could offer important constraints on the critical 
chemistry, on dynamics, on the 
heating and the cooling processes that occur in 
the primordial gas before and during gravitational collapse of protostructure. 
The research of primordial lines with HERSCHEL \cite{herschel:2001} 
could open an important new field of cosmology: the cosmochemistry. 
\\
Theory is essential for many aspects of astrochemistry (or cosmochemistry). 
Chemical models 
require chemical rates and these are not always available from 
experimentalists. The calculations of the minimum energy pathway and dynamical 
calculations are crucial. This last point is particularly important 
for lithium chemistry. Recently Zaldarriaga \& Loeb \cite{Zaldarriaga:2001} 
explored the imprint of the resonant 6708 \AA    
line opacity of neutral lithium 
on the temperature anisotropies of the CMB at observed wavelengths of 
250-350 $\mu$m. They showed that the standard CMB anisotropies would be 
significantly modified in this wavelength band. The primordial chemistry and 
particularly $LiH$ could give same conclusions and important consequences 
on the temperature and polarization anisotropies.
\\
Very recently LoSecco et al. \cite{losecco:2001} argued that extragalatic 
cold molecular clouds could lead to a significant absorption on the CMB. They 
speculate that the use of very high resolution spectrometers on large 
aperture telescopes might facilitate a 1-2 order of magnitude improvements in 
the CMB temperature measurements at high redshifts. Such accurate observations 
would enable us to constrain the anisotropy, inhomeogeneity of the Universe 
and the proto-chemistry. 
\\
We have seen the possibility of fragmentation of proto-clouds by the 
cooling due to $H_2$ and $HD$ molecules. This process could 
lead to the formation of primordial massive stars, which could be a possible 
source of contamination in heavier elements at early epochs. Then, 
the gravitational collapse of following objects (galaxies...) could be 
strongly influenced by the existence of heavier elements such as $CO$, $CI$ 
or $HCN$. 
\\
Chakrabarti \& Chakrabarti \cite{Chakrabarti:2000} showed that a 
significant amount of adenine, a DNA base, may be produced during 
molecular collapse, 
through the $HCN$ addition. Recently Sorrell \cite{Sorrell:2001} outlines a 
theoretical model for the chemical manufacture of interstellar amino acids 
and sugars. This chemistry model explains the existence of both the amino 
acid glycine and the sugar glycolaldehyde; this last component was 
recently detected in millimetre-wave rotational transition emission 
from the star-forming cloud Sagittarius B2 \cite{Hollis:2000}. The formation 
of DNA bases could happen in the early history of the Universe. Pre-biotic 
molecules could have contamined the first objects and planets from the 
beginning \cite{Puy:2001b}... 
\\
We are living in a golden age of astronomy, new observations with 
instruments such as NGST \cite{ngst:2001} PLANCK\cite{planck:2001} and 
HERSCHEL \cite{herschel:2001}, will push forward the 
frontiers of our ignorance. As Herbst wrote: {\it Astrochemistry may not tell 
us much about the first three minutes, but ultimately it should tell us our 
place in the Universe}.     

{\bf Acknowledgments.}
\\
We are very grateful to Marco De Petris, Massimo Gervasi and Fernanda 
Luppinacci for the opportunity to attend this splendid workshop. The authors 
gratefully acknowledge Francesco Melchiorri, Roberto Maoli and Pierre 
Encrenaz for valuable discussions on this field. 
Part of the work of D. Puy has been supported by the {\it D$^r$ Tomalla} 
Foundation and the Swiss National Science Foundation.

\end{document}